\documentclass[twocolumn,showpacs,prb,superscriptaddress,a4paper]{revtex4}

\usepackage{graphicx}
\usepackage{amsmath}
\usepackage{amssymb}

\begin{document}

\title{Band structure of ZnO from resonant x-ray emission spectroscopy}

\author{A. R. H. Preston} \email{andrew.preston@gmail.com}
  \affiliation{The MacDiarmid Institute for Advanced Materials and Nanotechnology, New Zealand}
  \affiliation{School of Chemical and Physical Sciences, 
    Victoria University of Wellington, 
    PO Box 600, Wellington 6140, New Zealand}
\author{B. J. Ruck}
  \affiliation{The MacDiarmid Institute for Advanced Materials and Nanotechnology, New Zealand}
  \affiliation{School of Chemical and Physical Sciences, 
    Victoria University of Wellington, 
    PO Box 600, Wellington 6140, New Zealand}
\author{L. F. J. Piper}
  \affiliation{Department of Physics, Boston University, 590 Commonwealth Ave., Boston, MA 02215}
\author{A. DeMasi}
  \affiliation{Department of Physics, Boston University, 590 Commonwealth Ave., Boston, MA 02215}
\author{K.~E.~Smith}
  \affiliation{Department of Physics, Boston University, 590 Commonwealth Ave., Boston, MA 02215}
\author{A.~Schleife}
  \affiliation{Institut f\"{u}r Festk\"{o}rpertheorie und -optik, Friedrich-Schiller-Universit\"{a}t, Max-Wien-Platz 1, 07743 Jena, Germany}
\author{F.~Fuchs}
  \affiliation{Institut f\"{u}r Festk\"{o}rpertheorie und -optik, Friedrich-Schiller-Universit\"{a}t, Max-Wien-Platz 1, 07743 Jena, Germany}
\author{F. Bechstedt}
  \affiliation{Institut f\"{u}r Festk\"{o}rpertheorie und -optik, Friedrich-Schiller-Universit\"{a}t, Max-Wien-Platz 1, 07743 Jena, Germany}
\author{J. Chai}
  \affiliation{The MacDiarmid Institute for Advanced Materials and Nanotechnology, New Zealand}
  \affiliation{Department of Electrical and Computer Engineering, University of Canterbury, Christchurch 8140, New Zealand}
\author{S. M. Durbin}
  \affiliation{The MacDiarmid Institute for Advanced Materials and Nanotechnology, New Zealand}
  \affiliation{Department of Electrical and Computer Engineering, University of Canterbury, Christchurch 8140, New Zealand}

\begin{abstract}
Soft x-ray emission and absorption spectroscopy of the O K-edge are employed to investigate the electronic structure of wurtzite ZnO(0001). A quasiparticle band structure calculated within the \textit{GW} approximation agrees well with the data, most notably with the energetic location of the Zn~3$d$ -- O~2$p$ hybridized state and the anisotropy of the absorption spectra. Dispersion in the band structure is mapped using the coherent \textbf{k}-selective part of the resonant x-ray emission spectra. We show that a more extensive mapping of the bands is possible in the case of crystalline anisotropy such as that found in ZnO.
\end{abstract}

\date{\today}
\maketitle

\section{Introduction}
The wide band gap semiconductor zinc oxide (ZnO) is a material with much potential in future electronic devices.~\cite{Science315.1377}
Despite significant interest in its fundamental electronic properties,~\cite{JApplPhys.98.041301} few spectroscopic studies of the full band structure exist.  A small number of angle-resolved photoemission spectroscopy (ARPES) studies have been reported,~\cite{PhysRevB.26.3144, JPhysCondMat.17.1271, JApplPhys.98.041301} yielding some agreement with density functional theory-based (DFT) calculations and providing information on surface and defect states. However, ARPES is limited to probing occupied states, and charging effects, surface preparation and sample quality are serious issues that can limit the accuracy of results. For example, photoemission measurements have variously located the fully occupied Zn 3{\it{d}} semi-core levels at energies ranging from 7.5 to 8.8~eV below the Fermi level.~\cite{PhysRevLett.27.97, JPhysCondMat.17.1271, PhysRevB.26.3144, PhysRevB.5.2296,  PhysRevB.9.600, JApplPhys.98.041301}  The close proximity of the $3d$ level to the O 2{\it{p}}-derived valence band has an appreciable impact on band structure calculations.~\cite{PhysRevB.52.13975, PhysRevB.47.6971, PhysRevB.37.8958, PhysRevB.52.R14316, PhysRevB.73.245212, PhysRevB.74.045202}  As a result, there is a need for bulk sensitive measurements of the ZnO shallow core-level and valence band dispersion. 

Here, we present angular dependent resonant x-ray emission spectroscopy (RXES; also known as resonant inelastic x-ray scattering -- RIXS) of the O K-edge of wurtzite ZnO crystals, for comparison with hybrid DFT calculations and quasiparticle energies obtained within the \textit{GW} framework.
Earlier studies used x-ray absorption and emission spectroscopy (XAS and XES) to directly measure the O 2{\it{p}} partial density of states (PDOS) of nano-structured ZnO~\cite{PhysRevB.68.165104, PhysRevB.70.195325} and its anisotropy in the conduction band.~\cite{JPhysCondMat.17.235} 
We go further by exploiting the dipole- and \textbf{k}-selection rules of RXES to map the anisotropic valence band of single-crystal ZnO.

RXES band mapping has been reported in simple materials like graphite, BN, SiC, and more,\cite{PhysRevLett.74.1234,PhysRevLett.76.4054,JElecSpecRelPhe.110.335} and reports on more complicated compounds related to ZnO are appearing.\cite{PhysRevB.72.085221,PhysRevB.73.115212,PhysRevB.75.165207,PhysRevB.77.125204} 
However, to our knowledge there is no literature examining the K-edge electronic structure of a post transition metal oxide in the detail we present, or on any material using our image-based technique. Further, we show crystal anisotropy to be a useful tool, allowing a larger part of the Brillouin zone to be uniquely accessed via RXES. Finally, compelling experimental evidence for use of the final-state rule in RXES is presented.

\section{Experiment and Theory}
The sample, a 500~nm ZnO epilayer, was grown on epi-ready (0001) sapphire substrates by plasma-assisted MBE, as described previously.~\cite{JElecMat.35.1316} As the sample was part of a separate study on doping, approximately 0.01\%~Ag was incorporated during growth, resulting in a layer averaged electron concentration of $7.3\times 10^{16}$~cm$^{-3}$. 
The high crystalline quality of the film was confirmed by RBS under the channeling condition,~\cite{JElecMat.36.472} a streaky RHEED pattern, and by low temperature photoluminescence, which shows donor bound exciton peaks are dominant with up to three LO phonon replicas.

The x-ray spectroscopy was performed on the undulator beamline X1B at the National Synchrotron Light Source at Brookhaven National Laboratory, which is equipped with a spherical grating monochromator and a Nordgren-type emission spectrometer. The energy resolution over the O K-edge was approximately 0.20~eV and 0.37~eV for the XAS and RXES respectively. XAS were recorded in total electron yield mode.  The photon energy was calibrated using the O K-edge and Ti L-edges of rutile TiO$_2$ measured during the experiment. The XES was calibrated with the 2nd order L-edge of Zn from both a calibration metal sample and the ZnO sample.

Due to its wurtzite structure, the ZnO \textit{p}-projected PDOS is strongly anisotropic.~\cite{JPhysCondMat.17.235}
To measure the anisotropy XAS and RXES were recorded for light incident at 20$^{\circ}$ and 70$^{\circ}$ relative to the sample normal, referred to as normal and grazing geometry, respectively.  Dipole selection rules mean that XAS recorded in the normal (grazing) geometry is dominated by the contributions from O 2$p_{xy}$ (2$p_{z}$) orbitals (see inset, Fig.~\ref{Fig1}). The opposite is true for RXES as the emission spectrometer is oriented perpendicular to the incident light: x-rays incident normal (grazing) to the sample result in grazing (normal) RXES. Thus $p_{xy}$ RXES is analysed with $p_z$ XAS and vice versa. This correspondence has been observed for wurtzite GaN by Strocov \textit{et al.}.~\cite{PhysRevB.72.085221}

Electronic structure calculations were performed within the hybrid DFT (HDFT) framework using the HSE03 functional for exchange and correlation.~\cite{JChemPhys.118.8207}  Quasiparticle effects were taken into account by a subsequent \textit{GW} correction of the HSE03 eigenvalues using many body perturbation theory.  In the \textit{GW} calculations, the Coulomb potential was fully screened using the random phase approximation dielectric function based on the HSE03 eigenvalues and functions.~\cite{PhysRevB.74.035101}
It has been shown that the \textit{ab initio} combination of HDFT and \textit{GW} calculations gives excellent values for the fundamental gaps and $d$-band positions of many semiconductors.~\cite{PhysRevB.76.115109}.
At present calculating a fully \textbf{k}-resolved \textit{GW} band structure is too computationally intensive, so we compare XAS and XES to PDOS obtained from the \textit{GW} calculation, and RXES to a modified HDFT band structure.

\section{Results and Discussion}
\begin{figure}
    \centering{\includegraphics[width=8cm]{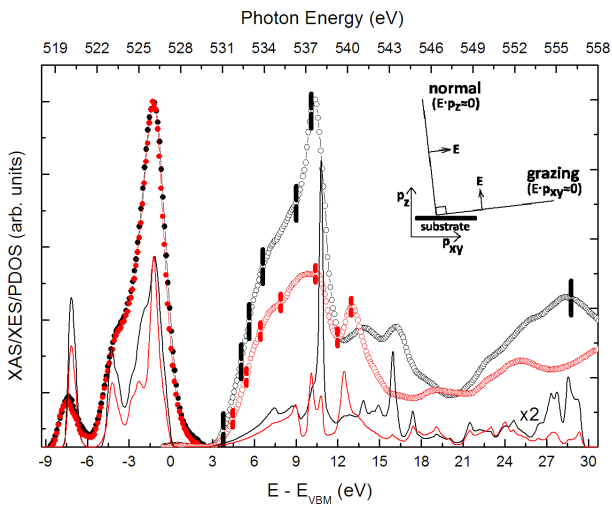}}
    \caption{(color online) $p_{xy}$ (black) and $p_z$ (red) orbital contributions to XES (closed symbols) and XAS (open symbols) compared to PDOS from the \textit{GW} calculation (lines). Bars indicate RXES excitation energies.}
    \label{Fig1}
\end{figure}

Figure~\ref{Fig1} shows the O K-edge XAS (open symbols) and XES, taken well above threshold (closed symbols). Strong linear dichroism associated with the anisotropy in the unoccupied states is seen in the XAS, especially in the first 10~eV above the conduction band minimum (CBM) where the $p_{xy}$ (black) contribution is considerably stronger than the $p_{z}$ (red) contribution. The XES is comprised of O 2$p$ states near the valence band maximum (VBM) and a clear peak corresponding to O 2$p$ states hybridized with Zn 3$d$ orbitals at 519.9~eV.

These spectra are compared with PDOS (solid lines) obtained from the \textit{GW} calculation. The PDOS have been convolved using gaussians with widths of 0.37~eV and 0.20~eV for the occupied and unoccupied states respectively. This is an estimate of the intrinsic instrumental resolution of the experiment; no lifetime effects have been considered. The PDOS have been aligned so that the main valence band peak matches with experiment. This yields excellent agreement with the entire XES, accurately locating the $3d$ hybrid. The weak anisotropy in the PDOS, combined with the lower resolution of the XES, explains the lack of contrast between $p_{xy}$ and $p_z$ XES.
The final state in XAS includes a core hole that is not included in the calculation, and which prevents absolute comparison between the energy scales of the XAS and XES.~\cite{PhysRevB.25.5150} Therefore, to compare to XAS, the theoretical unoccupied PDOS has been rigidly shifted by an additional $-1.0$~eV, providing an estimate of the core hole binding energy. With this adjustment the experimental agreement is excellent in terms of both the main peak locations and intensities, and the orbital anisotropy.
 
\begin{figure}
    \centering{\includegraphics[width=8cm]{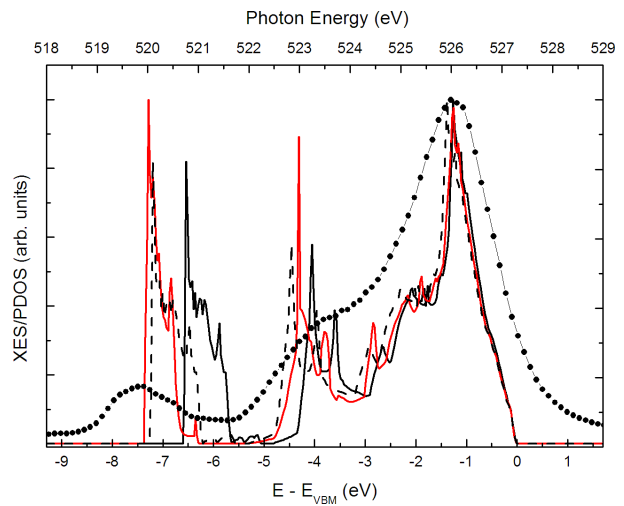}}
    \caption{(color online) Comparison between \textit{GW} (red line) and HDFT (solid black line) calculated O $p$-projected PDOS of the occupied states. The \textit{GW} calculation more accurately positions the Zn 3$d$--O 2$p$ hybrid when compared with experiment (closed symbols). Multiplying the HDFT energy scale by a factor of 1.1 results in improved agreement (dashed line).}
    \label{Fig2}
\end{figure}

The success of the \textit{GW} calculation is highlighted by a comparison with the HDFT result. We calculate a band gap of 3.2~eV, compared to 2.1~eV obtained from the HDFT calculation (and 3.4~eV experimentally.~\cite{JApplPhys.98.041301}) The valence band PDOS is also greatly improved: Figure~\ref{Fig2} shows the total O $2p$-projected occupied PDOS calculated using HDFT (black line), and with the \textit{GW} correction applied (red line). With \textit{GW} the Zn 3$d$ -- O 2$p$ hybrid peak is shifted to lower energy, peaking at $-7.1$~eV, in much closer agreement with the experimental peak at $-7.4$~eV (closed symbols), correcting a key flaw of standard DFT calculations.~\cite{PhysRevB.76.115109} With the Zn 3$d$ states located lower in energy the rest of the valence band is widened towards lower energies due to reduced $p$-$d$ repulsion. We note that scaling the HDFT energy axis by a factor of 1.1 is enough to obtain relatively close agreement with the \textit{GW} calculation (dashed line) for the occupied states.
An additional upward shift of 0.5~eV to the conduction band is sufficient to achieve similar agreement for the unoccupied states.
We have confirmed the accuracy of this approach approach by calculating the \textit{GW} band structure at selected \textbf{k} points.

\begin{figure}
    \centering{\includegraphics[width=8cm]{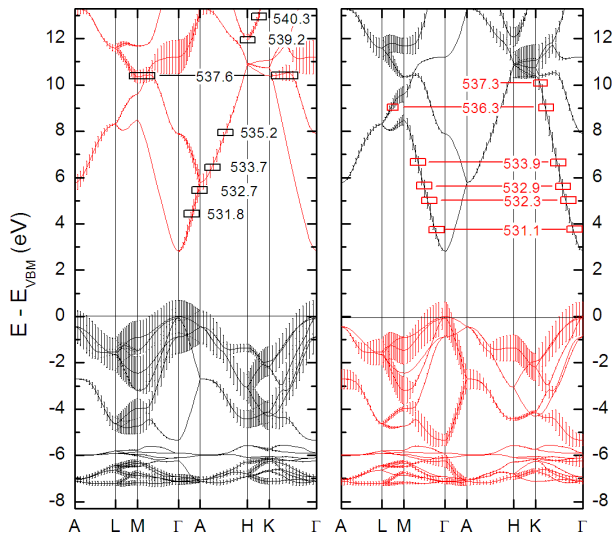}}
    \caption{(color online) HDFT band structure. Calculation is orbitally resolved into p$_{xy}$ (black) and p$_z$ (red) contributions. Error bars represent strength of contribution. The energy axis has been scaled for better agreement with \textit{GW} theory (see Fig.~\ref{Fig2}). Labels indicate RXES excitation energies.}
    \label{Fig3}
\end{figure}

Figure~\ref{Fig3} shows the band structure obtained from the HDFT calculation, with the above corrections applied. The $p_{xy}$ (black) and $p_z$ (red) orbital contributions to the eigenvalues are plotted independently; error bars represent the relative weight of the contribution. To simplify RXES analysis, the $p_{xy}$ ($p_z$) unoccupied states are plotted with $p_z$ ($p_{xy}$) occupied states (see above). RXES can be viewed as a scattering process where the final state includes an electron in the conduction band and a hole in the valence band, but no core-hole. So, unlike Fig.~\ref{Fig1}, no core hole correction is applied to the conduction band.

As expected, the band structure is that of a direct-gap semiconductor.
The strongly dispersing band that forms the CBM at $\Gamma$ is also anisotropic: only $p_{xy}$ states are found along $\Gamma$-M and $\Gamma$-K; $p_z$ along $\Gamma$-A-H.
This combination of strong dispersion and orbital selectivity make ZnO especially suitable for band mapping with RXES. For weakly correlated systems, and excitation energies close to the absorption edge, resonant effects contribute to XES, adding a \textbf{k}-conserving coherent contribution to the scattering. The crystal anisotropy, combined with orbital selection rules, adds further \textbf{k}-selectivity to the RXES. For ZnO, the use of these properties in concert allows valence band dispersion to be measured across L-M-$\Gamma$-A-H-K-$\Gamma$ of the Brillouin zone.

RXES were taken over a range of excitation energies, indicated by vertical bars on Fig.~\ref{Fig1}, and horizontal bars on Fig.~\ref{Fig3}.
The coherent fraction, extracted using standard techniques,~\cite{JElecSpecRelPhe.110.335} ranged from 0.45 to 0.6.

\begin{figure}
    \includegraphics[width=8cm]{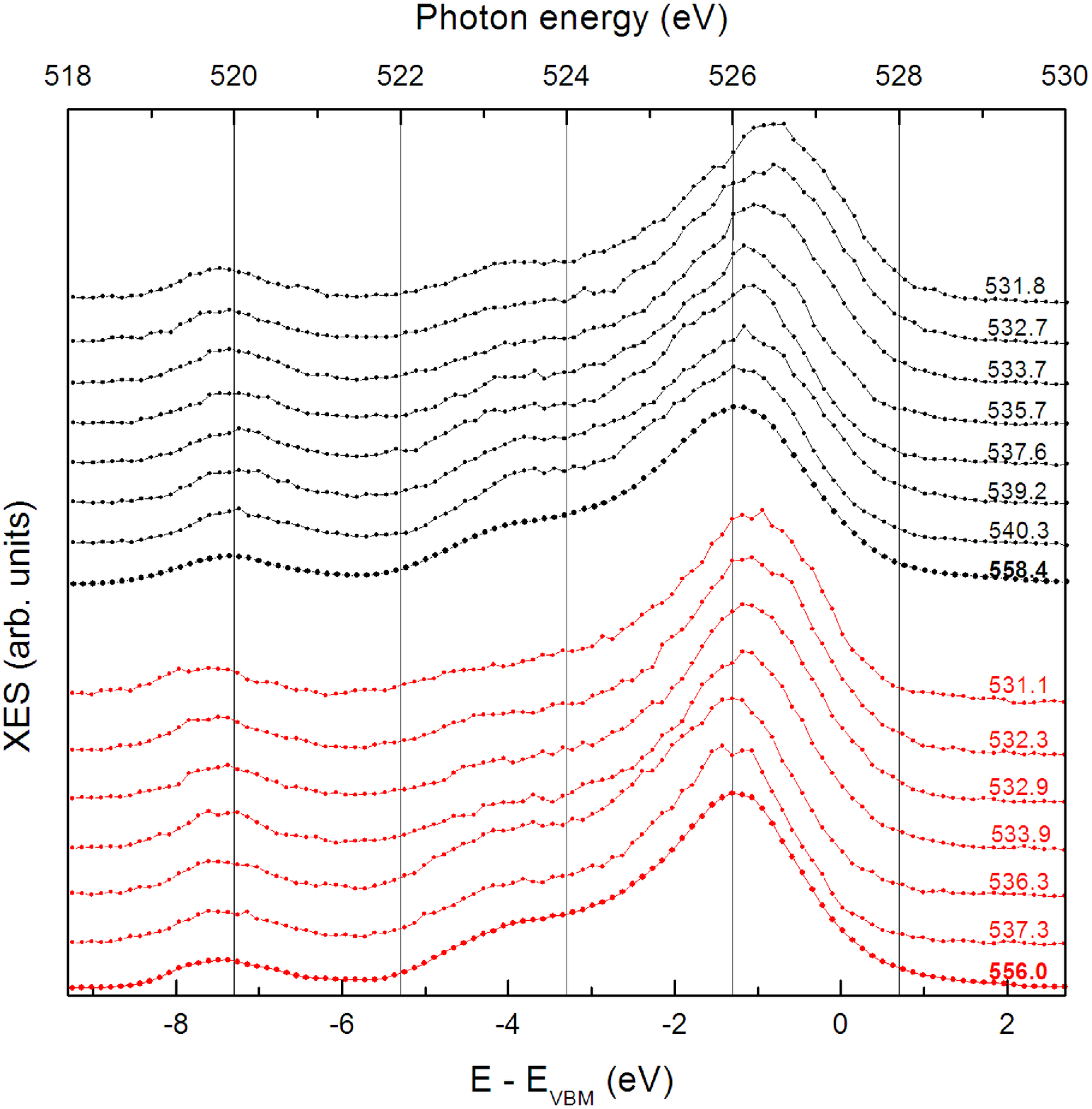}
    \caption{(color online) Normalized coherent RXES showing $p_{xy}$ (black) and p$_{z}$ (red) data. Labels indicate excitation energy ($h\nu$).}
    \label{Fig6}
\end{figure}

\begin{figure}
    \includegraphics[width=8cm]{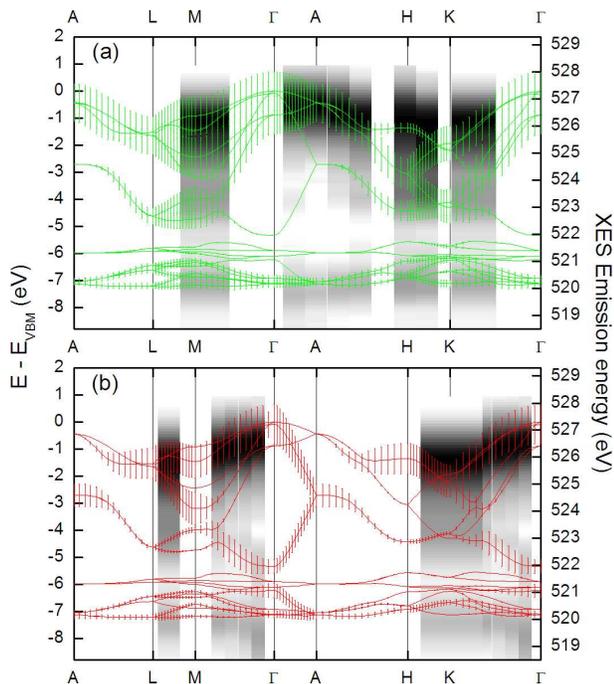}
    \caption{(color online) The coherent RXES overlaid with orbitally resolved band structure: (a)~p$_{xy}$ and (b)~p$_{z}$ orbitals. Intensity scales from zero (white) to one (black).}
    \label{Fig5}
\end{figure}

Using the above we construct Figure~\ref{Fig5}, which shows normalized intensity maps of the coherent fraction of each spectrum placed in \textbf{k}-space according to the position of the bars of Fig.~\ref{Fig3}. 
Figures~\ref{Fig5}(a) and (b) show the $p_{xy}$ and $p_z$ data, respectively. 
The imaging approach highlights the information contained in the raw spectra.
Dispersion toward higher energy at $\Gamma$, as expected for a direct-gap material,~\cite{JElecSpecRelPhe.110.335} is immediately obvious for both states.
For the $p_{xy}$ states dispersion along $\Gamma$-A-H-K is clearly resolved, with the upper band dispersing to lower energy. The bands around $-4$~eV decrease in energy and increase in intensity along A-H-K, in line with theory. The position of the Zn 3$d$ states is apparent, just below the theoretical bands. The spectrum obtained with an excitation energy of $537.6$~eV is located near both M and K, out of order with the other spectra, because of the strongly localised states in the conduction band at 10.5~eV.
The $p_z$ states add M-$\Gamma$ and K-$\Gamma$ to the parts of the brillioun zone we can resolve. The lower energy band near $-4$~eV disperses to $-5.5$~eV approaching $\Gamma$, in direct contrast to the observed behaviour of the $p_{xy}$ band in this energy range, and in agreement with theory.

There has been much discussion about the correct use of intermediate and final states in RXES calculations.~\cite{PhysRevB.59.7433,PhysRevB.73.115212,JElecSpecRelPhe.110.335} We note that if the core-hole correction were applied to the unoccupied states of Fig.~\ref{Fig3} (a downward shift of the theory by $1.0$~eV) the $p_{xy}$ spectrum taken with $537.6$~eV would be relocated to near $\Gamma$ and the strong intensity around $-4.5$~eV would be anomalous. This is clear evidence for the use of the final-state rule in RXES.

\section{Conclusion}
In conclusion we have used the strong anisotropy and dispersion of wurtzite ZnO, the \textbf{k}-selectivity of the RXES technique, and an HDFT+\textit{GW} calculation to construct a band mapping of ZnO across a wide range of high symmetry points.
The energy of the Zn 3$d$ core level is located and evidence for the use of the final state rule in RXES is presented.
It should be possible to use further high resolution RXES experiments to tune theoretical calculations across the technologically important post transition metal oxide and nitride series.

\begin{acknowledgments}
We acknowledge the assistance of R.~Mendelsberg and V.~J.~Kennedy.
Work at UOC was supported by the Marsden Fund (UOC0604) and the NZ TEC Doctoral Scholarship Program; at BU by the DOE (DE-FG02-98ER45680) and donors of the American Chemical Society Petroleum Research Fund; and at FSU by the European Community through NANOQUANTA (NMP4-CT-2004-500198), the Deutsche Forschungsgemeinschaft (BE1346/18-2 and /20-1) and the Carl-Zeiss-Stiftung.
The NSLS is supported by the DOE (DE-AC02-98CH10886). 
\end{acknowledgments}

\end{document}